\documentclass[a4paper]{jpconf}

\usepackage{iopams}  
\usepackage{graphicx}
\begin{document}

\title{Anomalous Conductance Quantization in the
Inter-band Gap of a One-dimensional Channel}

\author{Frederick Green$^1$ and Mukunda P. Das$^2$}

\address{$^1$ School of Physics, The University of New South Wales,
Sydney, NSW 2052, Australia.}
\address{$^2$ Department of Theoretical Physics,
RSPE, The Australian National University, Canberra, ACT 2601, Australia.}


\begin{abstract}
We report on a striking departure from the canonical step sequence of
quantized conductance in a ballistic, quasi-one-dimensional metallic channel.
Ideally, in such a structure, each sub-band population
contributes its Landauer conductance quantum independently of the rest.
In a picture based exclusively on coherent single-carrier transmission,
unitary back-scattering can lower a conductance step below ideal,
but it is absolutely impossible for it to enhance the ideal Landauer
conductance of a sub-band. Precisely such an anomalous and robust
nonlinear enhancement has already been observed over the whole
density range between sub-band thresholds
(de Picciotto R {\em et al} 2004 Phys. Rev. Lett. {\bf 92} 036805 and
2008 J. Phys. Condens. Matter {\bf 20} 164204).
We show theoretically that the anomalous enhancement of ideal Landauer
conductance is the hallmark of carrier transitions coupling
the discrete sub-bands.
\end{abstract}

\section{Introduction
}

Understanding the conductance in quantum-confined metallic channels
is a central aspect of electrical transport in meso- and
nanoscopic structures. Quantization of the conductance in
ballistic quantum-well channels is the unique property of their
one-dimensional (1D), waveguide-like nature.

The standard model of 1D quantized conductance \cite{FG,Davies,Yaro}
adopts the viewpoint of Landauer and associates \cite{RL,RLYI}
in which a highly constricted channel, interposed between
macroscopic ohmic contacts, is conceived as a simple
barrier potential modifying the propagation of
free single-electron quantum states.
Strong lateral confinement of these states by the device's
quantum-well structure leads to their segregation
into discrete levels (sub-bands),
separated by energy gaps whose magnitude may run from tens of meV
in III-V heterojunctions to several eV in carbon nanotubes.

This picture of conductance as quantum transmission accounts
elegantly for the well-documented resolution of the ohmic conductance
in 1D structures into a sequence of integral plateaux. As a function
of increasing carrier density in the device, each
successive step in the conductance extends, unaltered,
throughout the energy-gap region separating the discrete sub-bands.
A new plateau appears as soon as the chemical potential
and thus the carrier population cross the gap to access
the next higher sub-band.

That the single-particle quantum transmission picture does
not address all experimental observations is known
\cite{Neder,GTD,dP1,dP2}.
The reasons that it does not are also known
\cite{wims}.
First, quantum
transmission theory is restricted to weak-field linear response;
second, it does not account for the resistive dissipation that is
inevitable in every ohmic structure; third, the approach has
no way to address scattering processes other than purely coherent,
purely elastic, back-scattering.
These and other fundamental drawbacks of the transmission approach
have been critiqued in detail elsewhere
\cite{DG, DG2, DG3}.

One example of a crucial inelastic physical process is intra-band
scattering by phonon emission, responsible for the above-mentioned
dissipation. Another example is inter-band transitions, which
are inherently inelastic two-body processes with substantial energy exchange
between discrete bands. They lie beyond the scope of
a single-carrier description.

The physics of discrete transitions between distinct conduction
sub-bands lies at the heart of this paper. The experimental findings
that animate our work are those of de Picciotto,
Pfeiffer, Baldwin and West \cite{dP1,dP2}.

\vskip -6cm
\centerline{
\includegraphics[width=11truecm]{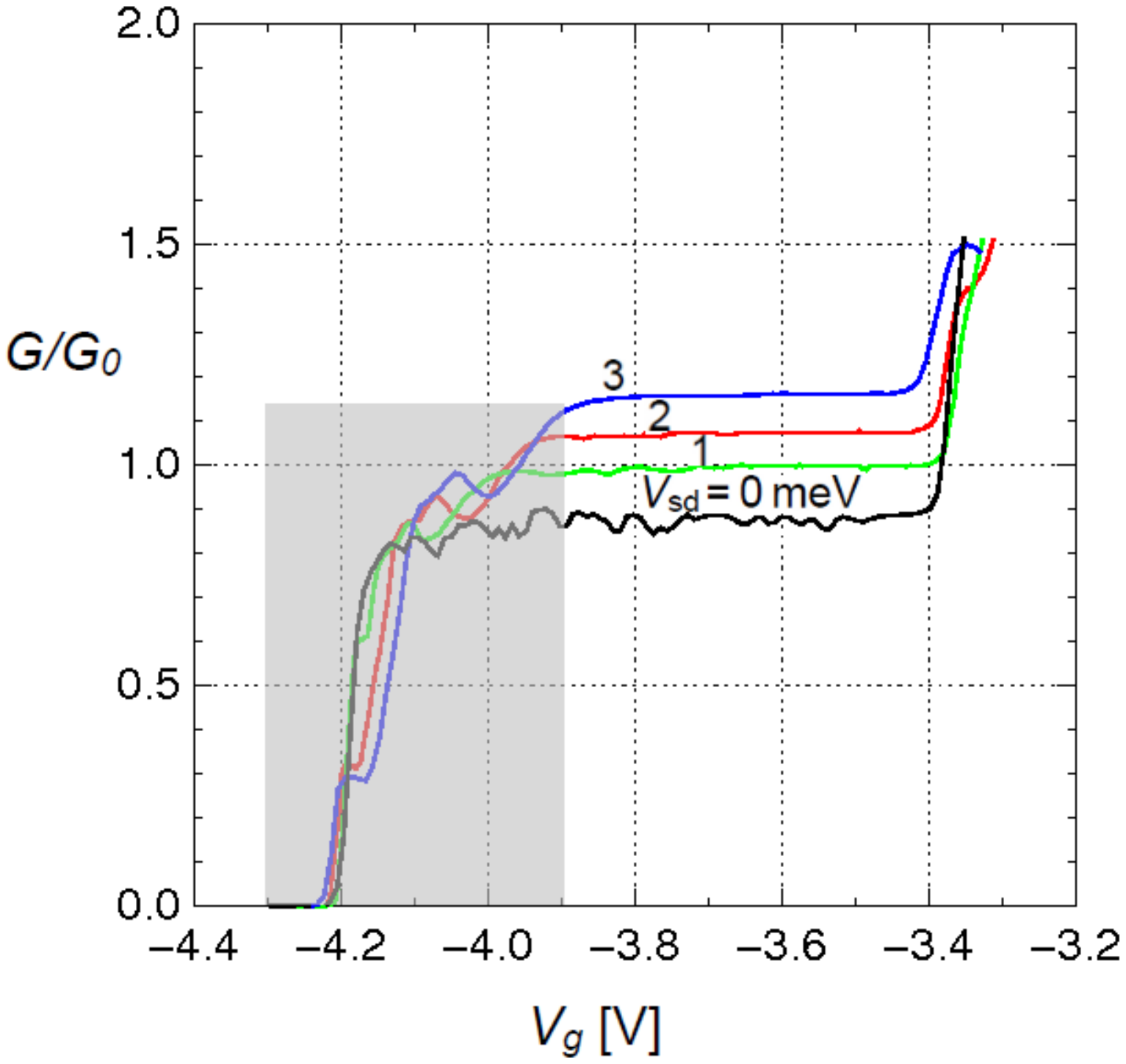}
}
{{\bf FIG. 1} {\small Quantized differential conductance measured in
a quasi-one-dimensional multi-sub-band ballistic channel,
after de Picciotto {\em et al}, reference \cite{dP2}.
Conductance is plotted in Landauer units
$G_0=77.48\mu$S, as a function of a control-gate voltage
modulating the total carrier density within the channel.
Beyond the labile structures (grey box) at threshold of the
ground-state band lies a series of very flat extended plateaux observed
at different values of source-drain bias voltage inducing the current.  
The system is taken beyond linear response, as seen in the
bias dependence of the conductance. Most remarkable is the raising
of $G$ above the absolute upper limit $G_0$ imposed by unitarity. This
shows that effects beyond simple quantum-coherent transmission
dominate the transport physics. 
}}
\vskip 0.25cm

Let us explain the importance of the remarkable non-linear,
non-Landauer 1D conductance plateaux documented by
de Picciotto and colleagues.
Figure 1 reproduces the core results of their references \cite{dP1}
and \cite{dP2}.
It shows a series of differential-conductance traces (normalized to the
Landauer quantum $G_0=77.48\mu$S) for a nearly
ideal ballistic quantum wire, taken at fixed source-drain driving
voltage and plotted as functions of gate voltage
controlling the channel chemical potential and so its carrier density.
The greyed region contains a complex of highly mutable shoulder
structures evident at the ground-state threshold, popularly
termed the ``0.7 anomaly'', which is not of interest here.

We focus, by contrast, upon the conductance plateaux extending over the
larger part of the inter-band region up to the threshold of the first
excited sub-band. Unlike the ``0.7'' features \cite{DG4},
they are robustly regular with a highly systematic dependence on the 
source-drain driving voltage.

\begin{itemize}
\item The anomalous steps are extremely flat and extend, with
carrier density, from the early-onset ``0.7'' feature sequence
right up to the threshold of the next higher sub-band.
Qualitatively and quantitatively, they are wholly distinct from
the relatively ephemeral structures close to first threshold.

\item The anomalous conductance steps are voltage dependent.
They are beyond any linear-response description.

\item These plateaux cover the entire region where the chemical
potential of the carriers lies in the gap between ground- and
first-excited-state bands. According to quantum-transmission theory,
such a structure cannot be higher than the ideal limit $G_0$.

\item With increasing source-drain voltage, the enhanced steps
increase in size from the expected weak-field baseline,
exceeding appreciably the absolute maximum set by $G_0$.
We stress the  {\em impossibility} of such a scenario within
quantum transmission, which would otherwise see its unitarity
(conservation of probability) wrecked.

\end{itemize}

The devices studied in references \cite{dP1} and \cite{dP2}
are of unprecedented quality,
perhaps the closest-to-perfect ballistic wires so far fabricated.
We remark on the care taken by the cited authors to isolate the
physics at work within their samples. It is a little surprising, then,
that (to our knowledge) no authors since the original team \cite{dP2} 
have commented on the startling violation of accepted predictions
by the quantized
conductance data of figure 1.

Reference \cite{dP1} contains a prescient comment
on the role of inter-mode coupling within the test
structures; that is, that there should be some
exchange of energy and possibly carriers between sub-bands,
setting up a mutual dynamical feedback.
To date, theoretical support for that hypothesis has not been at hand.
The goal is to provide it.
 
In the following we present a brief description of our quantum
kinetic analysis of the anomalous steps in 1D conductance.
Our microscopically conserving model is based on the quantum
Boltzmann formalism \cite{wims}
extended to inter-band transitions.
After this short account we survey our numerical results,
comparing and contrasting them with the basic data of
de Picciotto and colleagues. Particular features of
the results shed light on the relevant physics.
The paper ends with a summary and foreshadows novel
theoretical possibilities that could be tested in a renewed
series of experiments in similar high-quality structures.

\section{Problem and Solution}

The problem is to try to replicate and thus unpack the physics of
anomalous non-linear enhancement of the Landauer conductance.
We posit a uniform one-dimensional ballistic channel.

Since, even in principle, the active 1D device region cannot
be divorced from the large source and drain boundary leads,
its operative length is no longer exclusively determined by its physical
dimensions or the bulk mean free paths of its originating substrate.
Rather, its length is dictated by the longest carrier mean free path (MFP)
for the channel {\em as embedded in its non-ideal bounding leads}.
Adopting the estimate suggested in reference \cite{dP1}, we take
an operational channel length $L = 2 {\mu}$m. Our uniform-channel
results, however, do not depend on the absolute MFPs
assumed for this ballistic scenario.

In essence we are describing carrier behaviour averaged
over an abstract ensemble of such wires, seamlessly connected in series,
each with maximum mean free path $L$. At cryogenic temperatures
one expects the inelastic mean free path to set the longer scale,
fixing $L$. On the other hand, the observed weak-field conductance
falls short of the ideal Landauer quantum and we set the complementary
elastic intra-band MFP somewhat below $L$. An elastic
MFP of $0.769L$ matches the mean weak-field conductance
after reference \cite{dP1}.

\subsection{Transport equations with inter-band coupling}

Our transport equations define the behaviour
of two sub-band populations, separated by their energy gap $E_g$.
Given two steady-state distributions $f_{kj}$ for the lower band
($j=1$) and the next higher ($j=2$) as functions of band momentum
$k$, the equations are of modified Boltzmann-Drude form:

\begin{eqnarray}
{q{\cal E}\over \hbar} {\partial f_{k1}\over \partial k}
=&&
- R_{\rm in1}(f_{k1} - {\overline f}_{k1}) - R_{\rm el1} f^o_{k1}
\cr
&& {~~~ }
- R_{01} 
{\Bigl( e^{-E_g/k_{\rm B}T}  f_{k1} (1 - {\overline f}_{k2})
- {\overline f}_{k2} (1 - f_{k1}) \Bigr)},
\cr
{q{\cal E}\over \hbar} {\partial f_{k2}\over \partial k}
=&&
- R_{\rm in2} (f_{k2} - {\overline f}_{k2}) - R_{\rm el2} f^o_{k2} 
\cr
&& {~~~ }
- R_{02}
{\Bigl( f_{k2} (1 - {\overline f}_{k1}) -
e^{-E_g/k_{\rm B}T} ~{\overline f}_{k1} (1 - f_{k2}) \Bigr)}.
\label{nl01}
\end{eqnarray}

\noindent
The nature of the reference distributions ${\overline f}_{kj}$
is explained below.
Other notation is as follows. On the left-hand side of this pair of
steady-state equations, ${\cal E}$ is the uniform field exerted
on the carriers by the applied source-drain voltage.
On the right-hand sides the parameters
$R_{{\rm in}j}$ and $R_{{\rm el}j}$ are,
respectively, the intra-band inelastic and elastic scattering rates
assigned to each band while $R_{0j}$ correspondingly is the inter-band
transition rate for the coupling between the two populations.
The elastic collision term scales with the odd part of the distributions
since elastic scattering can only reverse the momentum direction
with no change in energy: $f^o_{kj} = (f_{kj} - f_{-kj})/2$.
 
The final parameter is the Boltzmann factor
$e^{-E_g/k_{\rm B}T}$ associated with promoting a carrier from
lower to upper sub-band across their gap separation $E_g$.
In the subsequent discussion we will express all
energies in thermal units $k_{\rm B}T$, and momenta in thermal units
$k_{\rm th} \equiv \sqrt{2k_{\rm B}T/\hbar^2}$.

We now discuss the meaning and role of the effective equilibrium functions

\begin{eqnarray}
&&
{~~~ ~~~ ~ }
{\overline f}_{k1}(\mu)
\equiv
1/(1 + \exp(k^2 - \mu)) ~~{\rm and}
\cr
\cr
&&{\overline f}_{k2}(\mu - E_g)
\equiv
1/(1 + \exp(k^2 + E_g - \mu))
\label{nl01.1}
\end{eqnarray}

\noindent
with momenta and energies in thermal units.
Transition events redistribute the electron population
between the two sub-bands of the channel.
The change in their respective densities
is presumed to depress the effective chemical potential of
the lower band: $\mu$ is renormalized to $\mu\!-\!\zeta_1$ while the
augmented population in the upper band, located above the lower
by the band gap $E_g$, follows the rise in its
effective chemical potential:
$\mu\!-\!E_g$ goes to $\mu\!+\!\zeta_2\!-\!E_g$.
Thus both quantities $\zeta_1$ and $\zeta_2$ should be non-negative.

For any choice of the pair of renormalized chemical potentials,
the coupled equations (\ref{nl01}) are solved systematically with
the expressions for ${\overline f}_{k1}(\mu\!-\!\zeta_1)$
and ${\overline f}_{k2}(\mu\!+\!\zeta_2\!-\!E_g)$,
equations (\ref{nl01.1}), as input.
Details of the relevant Green-function algorithm
are left to a longer account.

\subsection{Microscopic conservation}

Consider the equilibrium state at zero field, for which
$\zeta_1 = 0 = \zeta_2$. The effective equilibria are now
absolute and furthermore $f_{kj} = {\overline f}_{kj}$.
The distinct terms on either side of the transport equations (\ref{nl01})
all vanish individually. Detailed balance is satisfied.

What happens at finite field? The intra- and inter-band inelastic
collision terms on the right-hand sides of equations (\ref{nl01})
are not guaranteed to vanish identically when integrated separately,
although the left-hand-side expressions always do so.
As a consequence conservation must be imposed explicitly
upon the solution to the joint response.
\cite{DG,Greene,Mermin}
We have a type of generation-recombination problem; it follows
that conservation cannot apply to each sub-band individually,
but only jointly.

The bi-linear coupling between distributions in
the inter-band transition terms means that the full problem
cannot be solved even semi-analytically save in the trivial case
of independent sub-bands (zero transitions, $R_{01} = R_{02} = 0$).
Conservation then appears as a mandatory relation
linking the pair of potential shifts $\zeta_j$
so the system's total density is invariant
(the spin factor appears explicitly in the integrals,
which are rendered in dimensionless units):

\begin{eqnarray}
&&
2\int dk {\Bigl( {\overline f}_{k1}(\mu) +
{\overline f}_{k2}(\mu - E_g) \Bigr)}
\equiv
n(\mu)
\cr
&& {~~~ }
\equiv
2\int dk {\Bigl( f_{k1}(\mu - \zeta_1; \zeta_2) +
f_{k2}(\mu + \zeta_2 - E_g; \zeta_1) \Bigr)}. 
\label{nl02}
\end{eqnarray}

Since there are two undetermined quantities to solve,
a second relation is needed.
The new physical information to be adduced
must be independent of anything contained
in the transport equations themselves.

For the second, constitutive relation we take the Helmholtz
free-energy density for the non-equilibrium carrier distribution and
remove from it the formal energy of assembly for the system,
mediated by the chemical potential. Under the action of the
external driving field, the change in this net energy measures
the internal dynamical rearrangement of the sub-band distributions
induced by the field alone. This is manifested in the
renormalization of the bands' chemical potentials
$\mu_1\!\equiv\!\mu\!-\!\zeta_1 ~{\rm and}~
 \mu_2\!\equiv\!\mu\!+\!\zeta_2\!-\!E_g$,
as well as the form of the distributions $f_1(\mu_1)$ and $f_2(\mu_2)$.

Extending the standard thermodynamic expression \cite{TD}
for the Helmholtz free energy in each band,
we write its difference with the energy of assembly as

\begin{eqnarray}
&&F[f_j(\mu_j)]
\equiv
2\int dk (k^2\!-\!\mu_j) f_{kj}
\cr
&& {~~~ ~~~ ~~~ ~~~ ~~~ }
+ 2\!\! \int dk
{\Bigl( f_{kj}\ln f_{kj}\!+\!(\!1-\!f_{kj})\ln(1\!-\!f_{kj}) \Bigr)}
\label{nl03}
\end{eqnarray}

\noindent
recognizing the leading right-hand integral in equation (\ref{nl03})
as the total internal energy, less the assembly energy.
The second right-hand integral is the Uehling-Uhlenbeck entropy
entering into the H-theorem for fermions \cite{UU}.
At global equilibrium, minimizing $F$ (understood
as a functional of the distribution $f_j$ and subject to the latter's
variation) leads to the familiar Fermi-Dirac form for $f_j$.

We recapitulate.

\begin{itemize}

\item
A given gate voltage fixes the global chemical
potential and total density within the channel.

\item
The source-drain field, acting independently of this, excites
the carriers so a portion from the lower band is
promoted to the upper band.

\item
The density decrease in the lower band is determined by
the decrement $\zeta_1$ in the value of the common chemical
potential $\mu$, while
the increase in the upper band is determined by the increment
$\zeta_2$ in $\mu$.

\item
Any loss from the lower band must match the gain in the upper one.
Thus the two shifts in chemical potential are
coupled by the conservation relation (\ref{nl02}).

\item
To close the self-consistent solution for the $\zeta_j$,
we look for any change in the free energy of the system
as defined in equation (\ref{nl03}),
induced solely by application of the driving field.
\end{itemize}

The behaviour of $F$ above, purely as a function of chemical potential,
is quite different from its behaviour purely as a functional of $f$.
In the equilibrium state it is readily seen that

\[
{dF\over d\mu}[{\overline f}(\mu)] = -2\int dk {\overline f}_k(\mu)
\]

\noindent
so $F$ has no lower bound with increasing sub-band density.
Since the form of the $f_j$ is strictly prescribed
by solving the kinetic equations, the behaviour of $F$
is strictly a function of  $\mu_1$ and $\mu_2$.
The total net energy of the system, based on equation (\ref{nl03}),
must then exhibit maxima in the space $(\zeta_1,\zeta_2)$.

As our working principle we look for the maximum
in the total net energy summed over both bands,
as a function of the $\zeta_j$.
The physical picture is analogous to
a tap continuously feeding fluid to a finite container,
which finally overflows.
In a similar way the driven system will accumulate
as much of the inflowing excess energy as it can,
up to the point that increasing resistive dissipation
matches the inflow and precludes any further internal buildup.

Next we construct the difference of the sum
of net energies over the non-equilibrium distributions,
indexed by $\zeta_1$ and $\zeta_2$, and
its analogous non-equilibrium sum with $\zeta_2$ set to zero.
This difference vanishes for $\zeta_2 = 0$. Otherwise,
according to circumstances, it may exhibit a nontrivial maximum
in parameter space as $\zeta_2$ is increased systematically.
The computed quantity is

\begin{equation}
\Delta F(\mu; \zeta_1, \zeta_2)
= \sum_{j=1,2} {\Bigl(
F[f_j(\mu_j)] - {\Bigl. F[f_j(\mu_j)] \Bigr|}_{\zeta_2=0} \Bigr)}. 
\label{nl03.1}
\end{equation}

\noindent
Wherever we find the local maximum of $\Delta F$ on the locus
of constant density $n(\mu)$ there is a unique
self-consistent pair $(\zeta_1, \zeta_2)$.
This fixes the desired physical solution for the interacting system.
If no nontrivial solution exists, the maximum defaults to the
$\zeta_1$-axis (that is, $\zeta_2=0$)
and produces the standard Landauer conductance.

\vskip 0.25cm
\centerline{
\includegraphics[width=10truecm]{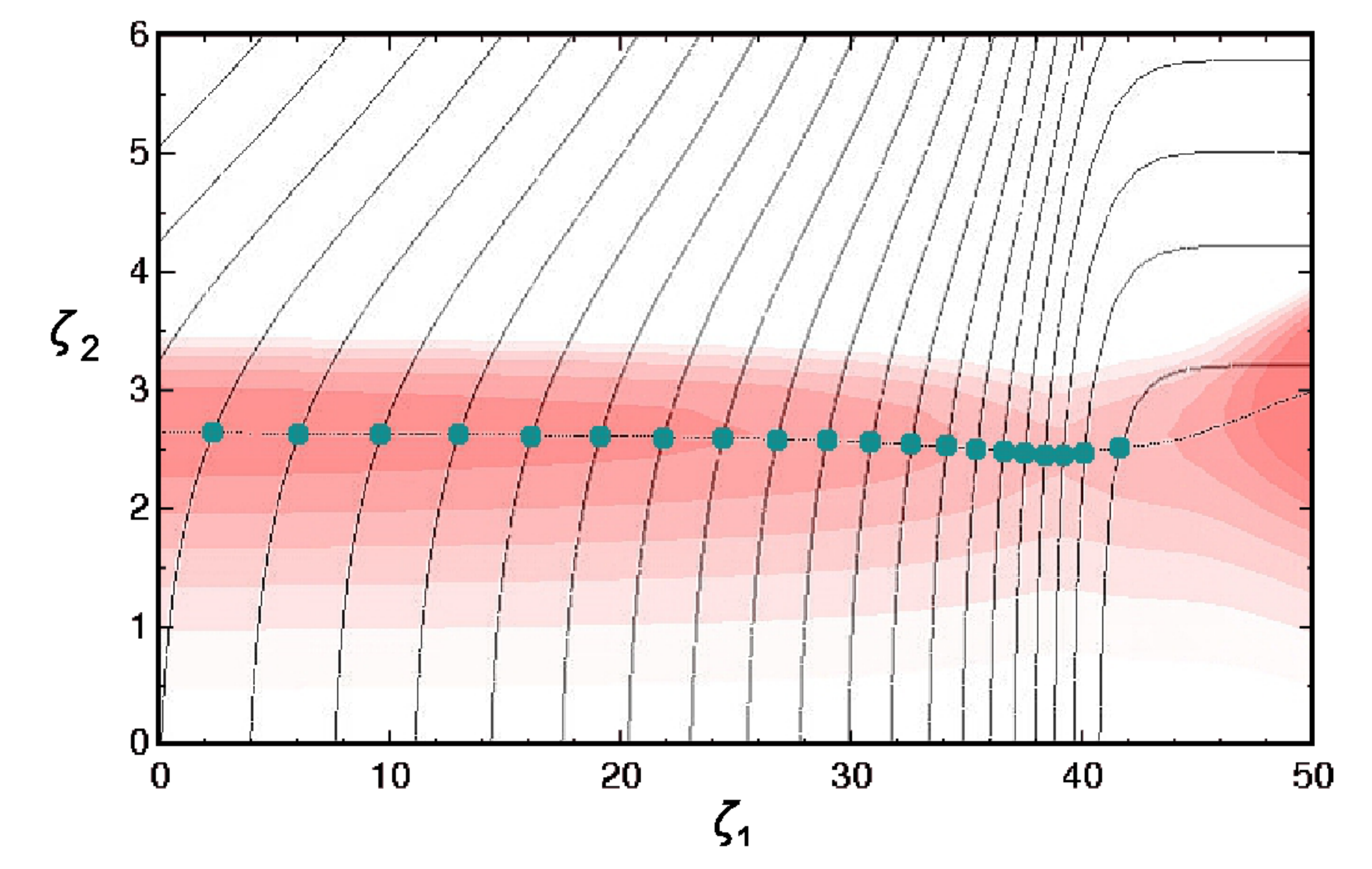}
}
{{\bf FIG. 2} {\small Landscape of self-consistent solutions
as functions of $\zeta_1$ and $\zeta_2$ (in thermal units) at 1mV
source-drain voltage over a channel 2$\mu$m long, at temperature 4K.
The dotted horizontal line at $\zeta_2 \approx 2.6$,
tracks the maximum in the bias-induced excess free energy 
$\Delta F(\mu; \zeta_1, \zeta_2)$, equation (\ref{nl03.1}),
over the latter's density map.
At global chemical potential $\mu$ the contours
of constant channel density $n(\mu)$, equation (\ref{nl02}),
rise vertically from the $\zeta_1$-axis to intersect the
maximum $\Delta F$ at the points of self-consistency (dots).
For each fixed density the solution determines
the system's non-linear conductance as a function of $\mu$.
Since the physical chemical potential is
$\mu = \mu_0 - \zeta'_1$ at the point of departure $\zeta'_1$ on
the $\zeta_1$-axis (with $\zeta'_1 = 0$ at origin),
the self-consistent values $(\zeta_1, \zeta_2)$
referred to $\mu_0$ are offset to yield
the physical renormalizations $(\zeta_1-\zeta_1', \zeta_2+\zeta_1')$
referred to the physical $\mu$.
}}
\vskip 0.25cm

Figure 2 illustrates the typical landscape of non-equilibrium
net free energy and total electron density in the space of the
renormalized chemical potentials, encompassing the family of
self-consistent solutions for a range of densities down from
a given equilibrium starting value $n(\mu_0)$.
The contour of constant density that matches the equilibrium value
$n(\mu_0 - \zeta'_1)$, say, is superimposed on the contours
of constant excess energy $\Delta F$.
For $\zeta_2 = 0$ the upper band is practically empty so each contour
of constant density departs orthogonally from the $\zeta_1$-axis.
The landscape is mapped as $\zeta_2$ increases and
the density in the upper band goes from near-empty to degenerate.

\subsection{Implementation}

Equation (\ref{nl03.1}) is a measure of the non-equilibrium excess
energy built up in the system when the driving field
induces redistribution of the populations between sub-bands.
Computing the self-consistent solution requires us
to connect the intra-band MFPs, together of course
with the inter-band transition probability, to the rates
$R_{{\rm in}j}, R_{{\rm el}j}$ and $R_{0j}$ that parametrize
the collision terms in the coupled equations (\ref{nl01}).

It is assumed that the mean free paths are
common to each sub-band, though this need not be so more generally.
Let $\lambda_{\rm in}$ be the inelastic MFP and $\lambda_{\rm el}$
be the elastic MFP. (Recall that, by hypothesis, the operational
channel length is given by $L = \lambda_{\rm in}$.) Any equilibrium
distribution has its associated characteristic velocity

\[
{\overline v}(\mu) \equiv - {v_{\rm th}\over {\overline f}_0(\mu)}
\int^{\infty}_0 k dk {\partial {\overline f}_k\over \partial k}(\mu) 
= v_{\rm th} (1 + e^{-\mu}) \int^{\infty}_0 dk {\overline f}_k(\mu).
\]

\noindent
We have scaled out the thermal velocity
$v_{\rm th} = \hbar k_{\rm th}/m^*$ where $m^*$ is the electron
effective mass.
In the low-density classical limit, this becomes essentially $v_{\rm th}$
while in the high-density degenerate limit it is the Fermi velocity
$v_{\rm th}\sqrt{\mu}$.
The quantity ${\overline v}$ thus sets the typical velocity scale.
Accordingly we define the rates in dimensionless units
from the respective characteristic velocities:

\begin{eqnarray}
R_{{\rm in}j}
\equiv&&
{L\over v_{\rm th}}{{\overline v}(\mu_j)\over \lambda_{\rm in}};
~~~~~~~~~
R_{{\rm el}j}
\equiv
{L\over v_{\rm th}}{{\overline v}(\mu_j)\over \lambda_{\rm el}}
\label{nl04}
\end{eqnarray}

The transition rates represent a different physical mechanism
and are treated differently from the intra-band ones,
as a single dimensionless parameter

\begin{equation}
R_{0j} \equiv L/\lambda_0
\label{nl05}
\end{equation}

\noindent
scaling inversely with a nominal ``transition MFP'' $\lambda_0$ which,
however, is qualitatively distinct from intra-band MFPs.
Its value is an experimental unknown. Moreover
$\lambda_0$ is likely to depend strongly on device geometry
and electrostatics \cite{dP1}. For this work
we set it an order of magnitude larger than the operational length $L$.

The current response summed over both parabolic sub-bands is given by

\begin{eqnarray}
I(\mu, V_{\rm sd})
&\equiv&
qk_{\rm th}v_{\rm th}
\int k{dk\over \pi}  \sum_{j=1,2} f_{kj}(\mu_j)
\cr
&=&
V_{\rm sd} {q^2\over \pi\hbar}
\int kdk \sum_{j=1,2}
{\left(
{\hbar k_{\rm th}v_{\rm th}\over q{\cal E} L}
f^o_{kj}(\mu_j)
\right)}
\label{nl05.1}
\end{eqnarray}

\noindent
where the source-drain voltage is $V_{\rm sd} = {\cal E}L$,
and we note that the even distributions do not contribute.
From equation (\ref{nl05.1}) all the transport properties are derived.

\section{Results}

We come now to the consequences for the quantized conductance.
In figure 3 below, for a total channel current $I(\mu, V_{\rm sd})$ at
a series of fixed applied $V_{\rm sd}$, we show the computed
chord conductance
$G = I/V_{\rm sd}$ for a device conforming to the specifications
of reference \cite{dP1}. The dynamical scattering
parameters are those of the preceding section, namely: operational length
$L \!=\! \lambda_{\rm in} \!=\! 2\mu$m;
$\lambda_{\rm el} \!=\! 0.769L \!=\! 1.538\mu$m;
$\lambda_0 \!=\! 10L \!=\! 20\mu$m.

Figure 4 of reference \cite{dP2} gives some evidence
of thermal broadening of conductance at the sub-band thresholds
presumably from localized Joule heating. We compute
our curves at the nominal temperature 4K.
The energy-gap value $E_g = 15$meV and the effective mass
for GaAs are used. Finally, on the horizontal axis of our
figure 3 we have mapped the global chemical potential $\mu$ to values
of a corresponding gate-control voltage, using the
parameters provided by reference \cite{dP1}.

Our figure 3 should be compared directly with figure 1 as taken from
reference \cite{dP2}, figure 3(a).
Both in the real data of figure 1 and in our calculation,
the action of a substantial source-drain voltage driving the
current through the channel leads to a series of elevated conductance
plateaux which
\bigskip

(a) are inherently non-linear in origin,

(b) are extremely flat and robust,

(c) anomalously exceed the Landauer upper bound on $G$ and thus

(d) violate the unitary limit of linear-response transmission theory.
\bigskip

\noindent
The striking confluence of behaviours between experiment and
theory speaks for itself.

\vskip -6cm
\centerline{
\includegraphics[width=11truecm]{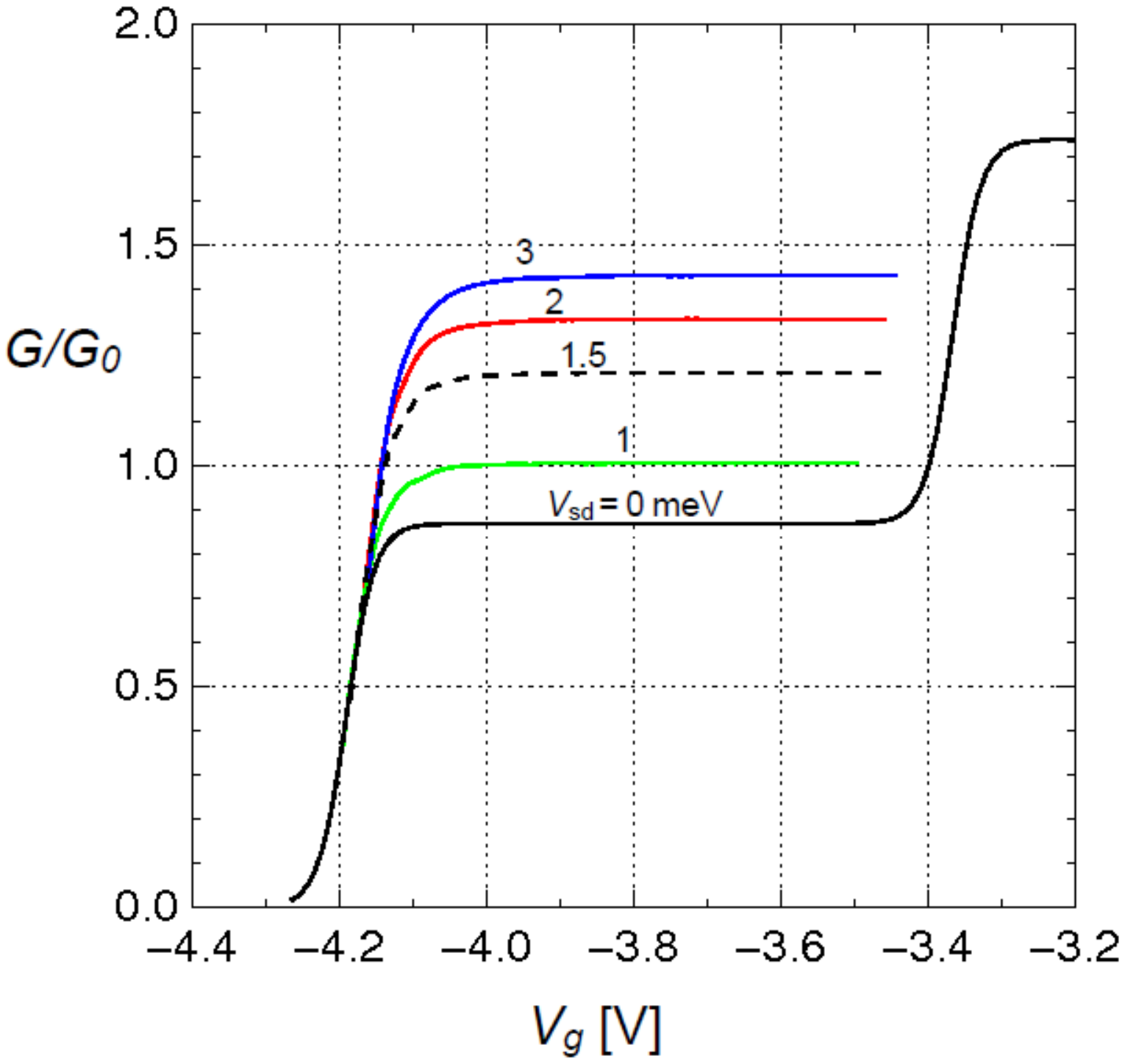}
}
\vskip 0.25cm
{{\bf FIG. 3} {\small Anomalous enhancement of conductance
$G$ calculated for a ballistic device equivalent to that of
figure 1 (figure 3(a) of reference \cite{dP2}) at a nominal
temperature 4K. Axis scales
are as for that figure; $G$ is plotted (units of $G_0 = q^2/\pi\hbar$)
versus gate voltage $V_g$
sweeping the channel density across the energy gap from the
bottom of the ground-state sub-band to the threshold of the first
excited-state sub-band. Bottom curve: in weak-field response
the quantized conductance matches
that for standard linear response. Higher curves: as the driving voltage
$V_{\rm sd}$ increases, the conductance acquires a non-linear enhancement.
The action of inter-band transitions dynamically redistributes
carrier density between sub-bands. This is responsible for the
strong enhancement of the step in $G$, beyond the unitarity limit
posited by quantum-transmission theories of conductance.
}}
\vskip 0.25cm

In figure 3(a) of reference \cite{dP2} (and figure 1
reproducing it in this paper) the plots show 
${\partial I/\partial V_{\rm sd}}$: the rate of change of current
with driving field, plotted as the density increases.
It is easily seen that when that slope is essentially flat over
a broad range of gate voltage (thus density) as in figure 1,
the simple conductance $I/V_{\rm sd}$, as in figure 3,
must track it closely and vice versa.

Before examining further characteristics of our theoretical conductance
we discuss differences between the present implementation
and the experiment. Our calculation here exhibits greater sensitivity
with respect to $V_{\rm sd}$ than the experiment so that, while in figure 3
the step increase of $G$ resulting from $V_{\rm sd} = 1$mV
coincides with that in figure 1, its height at 3mV is 1.43$G_0$
while its counterpart in figure 1 is 1.16$G_0$.
This overestimate might be accounted for in part
if the inelastic MFP $\lambda_{\rm in}$ suffered
shortening via optical-phonon emission at higher driving voltages
(the optical-phonon energy in GaAs is 35meV, not hugely
larger than $E_g$ at 15meV).
Use of energy-dependent mean free paths within
equations (\ref{nl01}), rather than fixed ones,
is an obvious aspect for exploration. Increased local
Joule heating with increased driving field may also suppress the
plateaux, as figure 4 suggests.

In figure 4 we show the properties of the enhanced conductance
taken at a typical mid-gap density where $G$ is steady.
Device specifications are the same as for figure 3, with an
additional choice of temperature, 8K as well as 4K.
Our calculated $G$ exhibits an onset at finite field
and asymptotic saturation at high fields. The threshold voltage value
at onset depends on temperature;
at low temperature the threshold value of $V_{\rm sd}$ tends to zero
and at high temperatures it rises in rough proportion to $T$.

\vskip -6cm
\centerline{
\includegraphics[width=11truecm]{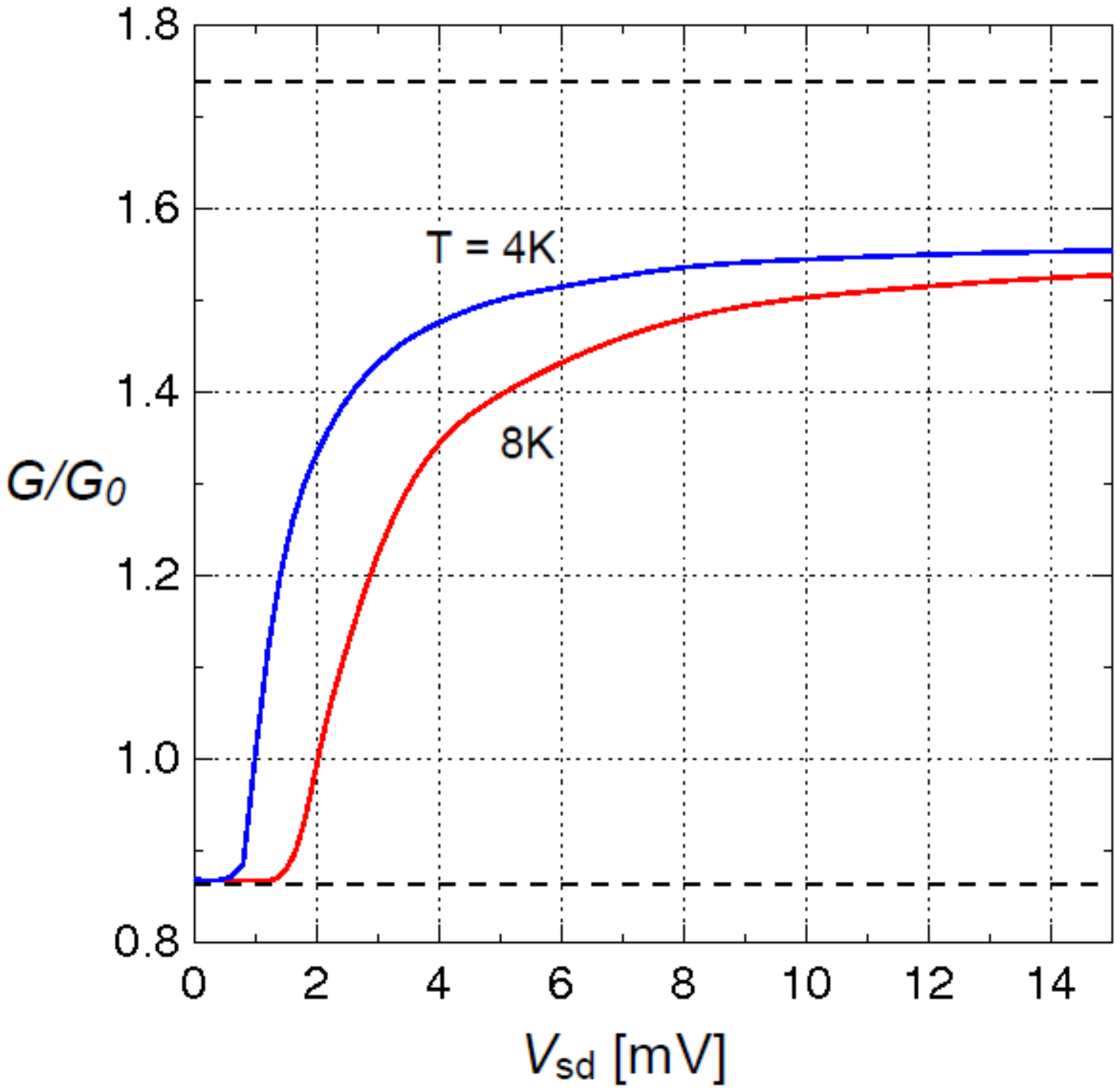}
}
\vskip 0.25cm
{{\bf FIG.4} {\small Threshold and saturation behaviour of
calculated conductance versus source-drain voltage, at temperatures
$T = 4$K and 8K. Lower dotted line is the first
Landauer level in this model, upper line is the second level.
The threshold driving field for onset of the enhancement scales
approximately with $T$. The saturation asymptote at high fields is
independent of temperature.}}
\vskip 0.25cm

The phenomenon above may partly explain why measurements
prior to de Picciotto {\em et al} \cite{dP1,dP2}
have not recorded the anomalous enhancement.
Predominantly, experiments in quantized conductance have been
carried out either at weak fields below threshold, or at higher
temperatures, or on noisy devices, or in any combination
of the above. Any enhancement of $G$ under
such conditions would tend to be washed out.

Saturation of $G$ in figure 4 sets in at driving fields considerably
higher than those employed in reference \cite{dP1} and in our
figure 3. The upper bound of $G$ is close to 1.6$G_0$, well short
of the second occupied level at 1.77$G_0$ in the weak-field limit.
This suggests that there is a field- and temperature-independent
maximum transfer of carriers to the upper band, beyond which
the feedback of transitions returning carriers
to the lower band precludes any increase. Saturation behaviour
may provide a further experimental window on the dynamics of
the inter-band transition.

\section{Summary and Implications}

De Picciotto {\em et al} \cite{dP1,dP2} in the first instance
addressed their experiments to the topic (still unresolved) of the
``0.7 anomaly''. Yet the same data harbours a message that is
perhaps more seminal to the understanding of ballistic transport
at low temperatures; namely, the quite surprising violation of the
unitary limit for quantized conductance.

That violation is illusory; the apparent paradox vanishes when
a more appropriate kinetic-theoretical argument is brought to bear,
going beyond the limits of single-particle transmission theory.
Carrier transitions between well separated sub-bands are
generation-recombination processes, viewed microscopically.
This means that their quantitative description must address
the direct creation and destruction of actual occupancies in such
discretely separate energy bands.
Unitarity still applies but, playing out as it must on the
much larger stage of multi-particle dynamics, it cannot
be accommodated by purely single-particle prescriptions.

Put succinctly, inelasticity and thermodynamic irreversibility
rule the physics.  It is crucial to build these into the theory
{\em explicitly}. That is not feasible within the restrictive
confines of reversible single-particle Hamiltonian dynamics.

De Picciotto and co-authors have presaged a role for inter-band
transitions in the dynamics of their structures \cite{dP1}. Our motive here
has been to advance a theory of such transitions in terms of
textbook quantum kinetics.

Certainly the boundary conditions for this problem do require
special care in interpretation to be given to the
ballistic nature of mesoscopic one-dimensional conductors.
Nevertheless it is unavoidable to confront a
transport problem where both elastic and inelastic scattering
processes act with equal physical
status, as in all normal metallic transport.

Our calculation strongly reinforces the conjecture \cite{dP1}
that inter-band transitions in a 1D ballistic device
do indeed produce anomalous
enhancement of the quantized conductance, within the density
regime between a sub-band threshold and its next-higher neighbour's.
This enhancement can indeed readily exceed the presumed unitarity
limit set by $G_0$, mandated as the absolute stepwise upper
bound for 1D conductance.

The straightforward reason for that phenomenon rests with
the physics of creation-annihilation across a band gap
and is not beholden to single-particle conservation band-by-band,
as it were in isolation. Only global conservation, subsuming
the bands within one interacting system, applies.
The central mechanism is non-linear feedback between
the coupled sub-band populations.

Finally, beyond explaining theoretically the quizzical enhancement
of ballistic conductance reported in references \cite{dP1} and \cite{dP2},
our results on temperature behaviour from figure 4 offer
a basis to predict thermal characteristics for
transition-induced changes in conductance in clean quantum wires.
Furthermore, there is a case for probing similar effects
in sufficiently clean carbon nanotubes, whose energy scale
and robustness at high fields far outstrip any device based
on GaAs heterojunction technology.
All of this would call for novel experiments.


\end{document}